\documentclass{article}
\usepackage{epsfig}
\def \be {\begin{equation}}
\def \ee {\end{equation}}
\def \bea {\begin{eqnarray}}
\def \eea {\end{eqnarray}}
\def \nn {\nonumber}

\def \a {\alpha}
\def \b {\beta}

\def \d {\delta}

\def \m {\mu}
\def \n {\nu}
\def \k {\kappa}

\def \s {\sigma}
\def \r {\rho}
\def \o {\omega}

\def \th {\theta}
\def \Th {\Theta}

\def \t {\tau}
\def \dag {\dagger}
\def \p {\partial}

\def\bd{\begin{document}}
\def\ed{\end{document}}
\def\nn{\nonumber}
\def\bea{\begin{eqnarray}}
\def\eea{\end{eqnarray}}
\let\bm=\bibitem
\let\la=\label

\def\N{{\cal N}}
\def\sst{\scriptscriptstyle}
\def\thetabar{\bar\theta}
\def\Tr{{\rm Tr}}
\def\one{\mbox{1 \kern-.59em {\rm l}}}

\def\a{\alpha}      \def\da{{\dot\alpha}}
\def\b{\beta}       \def\db{{\dot\beta}}
\def\c{\gamma}  \def\C{\Gamma}  \def\cdt{\dot\gamma}
\def\d{\delta}  \def\D{\Delta}  \def\ddt{\dot\delta}
\def\e{\epsilon}        \def\vare{\varepsilon}
\def\f{\phi}    \def\F{\Phi}    \def\vvf{\f}
\def\h{\eta}
\def\k{\kappa}
\def\l{\lambda} \def\L{\Lambda}
\def\m{\mu} \def\n{\nu}
\def\o{\omega}
\def\P{\Pi}
\def\r{\rho}
\def\s{\sigma}  \def\S{\Sigma}
\def\t{\tau}
\def\th{\theta} \def\Th{\Theta} \def\vth{\vartheta}
\def\X{\Xeta}
\def\z{\zeta}
\def\w{\wedge}
\def\u{\underline}
\def\hs{\hspace}

\def\cA{{\cal A}} \def\cB{{\cal B}} \def\cC{{\cal C}}
\def\cD{{\cal D}} \def\cE{{\cal E}} \def\cF{{\cal F}}
\def\cG{{\cal G}} \def\cH{{\cal H}} \def\cI{{\cal I}}
\def\cJ{{\cal J}} \def\cK{{\cal K}} \def\cL{{\cal L}}
\def\cM{{\cal M}} \def\cN{{\cal N}} \def\cO{{\cal O}}
\def\cP{{\cal P}} \def\cQ{{\cal Q}} \def\cR{{\cal R}}
\def\cS{{\cal S}} \def\cT{{\cal T}} \def\cU{{\cal U}}
\def\cV{{\cal V}} \def\cW{{\cal W}} \def\cX{{\cal X}}
\def\cY{{\cal Y}} \def\cZ{{\cal Z}}

\def\ua{\underline{\alpha}} \def\ubb{\underline{\beta}}
\def\ug{\underline{\gamma}}
\def\ub{\underline{\phantom{\alpha}}\!\!\!\beta}
\def\uc{\underline{\phantom{\alpha}}\!\!\!\gamma}
\def\um{\underline{\mu}} \def\un{\underline{\nu}}
\def\ud{\underline\delta}
\def\ue{\underline\epsilon}
\def\una{\underline a}\def\unA{\underline A}
\def\unb{\underline b}\def\unB{\underline B}
\def\unc{\underline c}\def\unC{\underline C}
\def\und{\underline d}\def\unD{\underline D}
\def\une{\underline e}\def\unE{\underline E}
\def\unf{\underline{\phantom{e}}\!\!\!\! f}\def\unF{\underline F}
\def\unm{\underline m}\def\unM{\underline M}
\def\unn{\underline n}\def\unN{\underline N}
\def\unp{\underline{\phantom{a}}\!\!\! p}\def\unP{\underline P}
\def\unq{\underline{\phantom{a}}\!\!\! q}
\def\unQ{\underline{\phantom{A}}\!\!\!\! Q}
\def\unH{\underline{H}}
\def\ul{\underline}

\def\As {{A \hspace{-6.4pt} \slash}\;}
\def\bs {{b \hspace{-6.4pt} \slash}\;}
\def\Ds {{D \hspace{-6.4pt} \slash}\;}
\def\ds {{\del \hspace{-6.4pt} \slash}\;}
\def\ss {{\s \hspace{-6.4pt} \slash}\;}
\def\ks {{ k \hspace{-6.4pt} \slash}\;}
\def\ps {{p \hspace{-6.4pt} \slash}\;}
\def\pas {{{p_1} \hspace{-6.4pt} \slash}\;}
\def\pbs {{{p_2} \hspace{-6.4pt} \slash}\;}

\def\Fh{\hat{F}}
\def\Vh{\hat{V}}
\def\Xh{\hat{X}}
\def\ah{\hat{a}}
\def\xh{\hat{x}}
\def\yh{\hat{y}}
\def\ph{\hat{p}}
\def\xih{\hat{\xi}}

\def\psit{\tilde{\psi}}
\def\Psit{\tilde{\Psi}}
\def\tht{\tilde{\th}}

\def\At{\tilde{A}}
\def\Qt{\tilde{Q}}
\def\Rt{\tilde{R}}
\def\Nt{\tilde{N}}

\def\at{\tilde{a}}
\def\st{\tilde{s}}
\def\ft{\tilde{f}}
\def\pt{\tilde{p}}
\def\qt{\tilde{q}}
\def\vt{\tilde{v}}
\def\nt{\tilde{n}}

\def\delb{\bar{\partial}}
\def\bz{\bar{z}}
\def\bD{\bar{D}}
\def\bB{\bar{B}}

\def\bk{{\bf k}}
\def\bl{{\bf l}}
\def\bp{{\bf p}}
\def\bq{{\bf q}}
\def\br{{\bf r}}
\def\bx{{\bf x}}
\def\by{{\bf y}}
\def\bR{{\bf R}}
\def\bV{{\bf V}}

\def\d{\delta}\def\D{\Delta}\def\ddt{\dot\delta}

\def\p{\partial} \def\del{\partial}
\def\xx{\times}
\def\uno{\mbox{1 \kern-.59em {\rm l}}}

\def\trp{^{\top}}
\def\inv{^{-1}}
\def\dag{{^{\dagger}}}

\def\pr{\prime}
\def\rar{\rightarrow}
\def\lar{\leftarrow}
\def\lrar{\leftrightarrow}

\def\cp{{\bf CP}^3}
\def\omg1{\omega_1}
\def\omg2{\omega_2}
\def\omg3{\omega_3}
\def\n1{n_1}
\def\n2{n_2}
\def\n3{n_3}
\def\J1{J_1}
\def\J2{J_2}
\def\J3{J_3}

\begin{document}

\title{Multi-spin strings in $AdS_4\times CP^3$ and its $\beta$-deformation}
\author{Jun-Bao Wu\footnote{E-mail: wujb@ihep.ac.cn}\\
\small{Institute of High Energy Physics,
and Theoretical Physics Center for Science Facilities,}\\
\small{ Chinese Academy of Sciences,
Beijing 100049, P.R. China}\\
}
\maketitle

\abstract{In this paper, we study the multi-spin string solutions in $AdS_4\times CP^3$ and its $\beta$-deformation with
real $\beta$. We give various explicit solutions after some general studies. Conserved charges are computed for these explicit
solutions.}

\section{Introduction}

Macroscopic classical string solutions with large angular momentums in $AdS_5\times S^5$ \cite{GKP2002}
are dual to composite gauge invariant operators with large quantum numbers in ${\cal N}=4$ super Yang-Mills theory.
We can perform semiclassical quantization around these string solutions and compute, for examples, the one-loop
corrections to the energy of these strings \cite{Frolov:2002av}. For certain region of the parameters, the energy of the classical string remarkably
matches with the anomalous dimension of the composite operator \cite{Beisert:2003xu, Beisert:2003ea}, computed using the spin chain technique \cite{Minahan:2002ve}. This match provides strong supports to the AdS/CFT correspondence \cite{Mal97, Gubser:1998bc, Witten:1998qj} beyond the supergravity limit (works on semiclassical string solutions in $AdS_5\times S^5$ were reviewed in \cite{Tseytlin:2010jv}).
Later, semi-classical strings and composite operators were studied using the algebraic curves approach \cite{Kazakov:2004qf}-\cite{Beisert:2005di} (for a recent review, see \cite{SchaferNameki:2010jy}).

Recently correspondence between three-dimension ${\cal N}=6$ Chern-Simons-matter theory and type IIA string theory on $AdS_4\times CP^3$ background was established in \cite{ABJM}. 
Integrability in this $AdS_4/CFT_3$ correspondence
\cite{Minahan:2008hf}-\cite{Kristjansen:2008ib}  was reviewed in
\cite{Klose:2010ki, Lipstein:2011}. Various explicit classical
string solutions were found in \cite{Chen:2008qq}, complementing the
studies using the algebraic curves in \cite{Gromov:2008bz} .
One-loop corrections to the energy of various classical strings were
computed in \cite{Astolfi:2008ji}-\cite{Beccaria:2012vb}. Splitting
of folded strings in \cite{Chen:2008qq} was studied in
\cite{Wu:2011rj}.

Examples of gauge/gravity correspondences with less supersymmetries
are certainly with general interests. One of the methods to get
conformal field theories with less supersymmetries, starting from
${\cal N}=4$ super Yang-Mills theory, is through marginal
deformations \cite{Leigh:1995ep}. Among these deformations, there is
so-called $\beta$-deformation which preserves ${\cal N}=1$
supersymmetry and simply adds certain phase factors to the
interaction terms in the Lagarangian. This deformation can be
elegantly written in term of some kind of star products
\cite{Lunin:2005jy}. This fact leads to the results that all planar
partial amplitudes are given by the product of corresponding ${\cal
N}=4$ amplitudes and an overall phase determined by the external
particles \cite{Khoze:2005nd}. This result at tree level was also
obtained \cite{Gao:2006mw} in the framework of twistor string theory
\cite{Witten:2003nn} by applying and generalizing the prescription
in \cite{Kulaxizi:2004pa}. Bethe ansatz for this theory was showed
to involve twisted boundary conditions
\cite{Roiban:2003dw}-\cite{Ahn:2010ws}. The gravity dual of this
deformed gauge theory was found in \cite{Lunin:2005jy}, using a
solution generating technique. Various classical string solutions in
this background were found \cite{Frolov:2005ty}-\cite{Ahn:2012hs}
which lead to some precise checks of this gauge/gravity
correspondence with less supersymmetries (for other discussions, we
refer to the review \cite{Zoubos:2010kh}). The understanding of this
deformation was improved in \cite{Imeroni:2008cr} where similar
deformation in ABJM theory was also studied in detail. Giant magnons
and single spike solutions in the deformed $AdS_4\times CP^3$ was
found in \cite{Schimpf:2009rk}. Finite-size effect on the dispersion
relation of giant magnons was studied in \cite{Ahn:2011nm}
\footnote{similar finite-size effect in undeformed case was studied
in \cite{Grignani:2008te}.}. $\beta$-deformed BLG
\cite{BL1}-\cite{G2} theory was studied in \cite{Berman:2008be}.
This marginal deformations in BLG theory and ABJM theory were also
studied in \cite{Akerblom:2009gx} using $3$-algebra. Similar
deformed solutions for membranes and fivebranes in eleven
dimensional supergravity ant their near horizon limits were given in
\cite{Berman:2007tf}.

The integrability in $\beta$-deformed ABJM theory and its gravity
dual certainly deserves further investigation.
In this paper, we will discuss  some multi-spin string solutions in
$AdS_4\times CP^3$ and its $\beta-$deformations
\footnote{Multi-spin strings in $AdS_5\times S^5$ were studied in \cite{Frolov:2003qc}.
One-loop correction to the energy of some special simple multi-spin strings in $AdS_4\times CP^3$
was studied in \cite{Bandres:2009kw}. We will study more general multi-spin string solutions in section 2. Some classical strings with several angular momenta, in $AdS_5\times S^5$ and $AdS_4\times CP^3$ were studied in  \cite{Giardino:2011dz, Giardino:2011uc}, while they are not belong to the class of multispin string solutions studied in this paper.}.

 In the next section, we give some special multi-spin solution in $AdS_4\times CP^3$ after some general discussions.
 In section~\ref{deformed},
 we focus on multi-spin string solutions in $\beta$-deformed $AdS_4\times CP^3$. After giving one special solution,
we study the solutions in deformed spacetime starting from certain ansatz in section~\ref{undeformed}.
We put the conclusion and some discussions in the final section.

\section{Multi-spin strings in $AdS_4\times CP^3$\label{undeformed}}
\subsection{General studies}
The metric on $AdS_4\times CP^3$ is
\be ds^2=R^2(\frac14ds^2_{AdS_4}+ds^2_{CP^3}),\ee
with
\be ds^2_{AdS_4}=-\cosh^2\rho dt^2+d\rho^2+\sinh^2\rho(d\theta^2+\sin^2\theta d\varphi^2),\ee
and \bea ds^2_{CP^3}&=&d\xi^2+\frac14\cos^2\xi(d\theta_1^2+\sin^2\theta_1
d\varphi_1^2)+\frac14\sin^2\xi(d\theta_2^2+\sin^2\theta_2 d\varphi_2^2))\nn\\
&&\,\,\,+\cos^2\xi\sin^2\xi
(d\psi+\frac12\cos\theta_1d\varphi_1-\frac12\cos\theta_2d\varphi_2)^2, \eea
where $0\le\xi<\frac\pi2,  -\pi\le\psi<\pi, 0\le\theta_i\le\pi,
0\le\varphi_i<2\pi$.
The relation between $R$ and the 't~Hooft coupling $\lambda=N/k$ in the gauge theory side is
\be R=2^{5/4}\pi^{1/2}\lambda^{1/4}\alpha^{\prime1/2}.\label{r}\ee
The corresponding three-dimensional superconformal theory is Chern-Simons-matter theory with
gauge group $U(N)_k\times U(N)_{-k}$. The matter superfields are bifundamental superfields $A_1, A_2$
and anti-bifundamental superfields $B_1, B_2$. The superpotential is
\be W=\frac{4\pi}{k}\Tr(A_1B_1A_2B_2-A_1B_2A_2B_1).\label{superpotential}\ee

Since the NS-NS B-field vanishes,
the $\sigma$-model action for the string in the conformal gauge is
\bea S&=&\frac{R^2}{4\pi\alpha^\prime}\int d\sigma d\tau \left(\frac14(\cosh^2\rho(\dot{t}^2-t^{\prime2})
-\dot{\rho}^2+\rho^{\prime2}+\sinh^2\rho(-\dot{\theta}^2+\theta^{\prime2})\right.\nn\\
&&\,\,+\sinh^2\rho\sin^2\theta(-\dot{\varphi}^2+\varphi^{\prime 2}))-\dot{\xi^2}+\xi^{\prime2}+\frac14\cos^2\xi(\theta_1^{\prime2}+\sin^2\theta_1\varphi^{\prime2}_1\nn\\
&&\,\,-(\dot{\theta}_1^2
+\sin^2\theta_1\dot{\varphi}_1^2))+\frac14\sin^2\xi(-(\dot{\theta}_2^2+\sin^2\theta_2\dot{\varphi}_2^2)
+\theta_2^{\prime2}+\sin^2\theta_2\varphi^{\prime2}_2)\nn\\
&&\,\,+\sin^2\xi\cos^2\xi \left(-(\dot{\psi}+\frac12\cos\theta_1\dot{\varphi}_1-\frac12\cos\theta_2\dot{\varphi}_2)^2
\right.\nn\\ &&\,\,\left.\left.
+(\psi^\prime+\frac12\cos\theta_1\varphi_1^\prime-\frac12\cos\theta_2\varphi_2^\prime)^2\right) \right).\eea
From eq.~(\ref{r}), we get
\be\frac{R^2}{4\pi\alpha^\prime}=\frac{\sqrt{32\pi^2\lambda}}{4\pi}.\ee
In the following, we introduce parameter $\tilde\lambda$ defined by $\tilde\lambda\equiv32\pi^2\lambda$.

We consider the following ansatz for the multi-spin string solutions
\be \rho=0, t=\kappa\tau, \psi=\omega_1\tau+n_1\sigma, \varphi_1=\omega_2\tau+n_2\sigma, \varphi_2=\omega_3\tau+n_3\sigma, \ee
with $\xi, \theta_1, \theta_2$ being constants. We notice that $n_2, n_3, \frac{n_2+n_3}2+n_1$ should be integers and this can be understood from eq.~(2.7-10)
in \cite{Chen:2008qq}.

For this ansatz it is not hard to see that the equations of motion for $t, \rho, \theta, \varphi, \psi,$ $\varphi_1,
\varphi_2$ are automatically satisfied.
So we only need to consider the equations of motions for $\xi, \theta_1$ and $\theta_2$:
\bea  \frac12\sin4\xi(\tilde\omega^2-
\tilde n^2)-\frac14\sin2\xi(\sin^2\theta_1(\omega_2^2-n_2^2)+
\sin^2\theta_2(\omega_3^2-n_3^2))=0, \eea
\be (-\omega_2\tilde\omega+n_2\tilde n)\sin\theta_1
 \sin^2\xi\cos^2\xi+\frac14\cos^2\xi\sin2\theta_1(\omega_2^2-n_2^2)=0,
\ee
\be (\omega_3\tilde\omega-n_3\tilde n)\sin\theta_2
 \sin^2\xi\cos^2\xi+\frac14\sin^2\xi\sin2\theta_2(\omega_3^2-n_3^2)=0,
\ee
where we have defined that
\bea \tilde\omega&\equiv&\omega_1+\frac{\omega_2}{2}\cos\theta_1-\frac{\omega_3}2\cos\theta_2,\\
\tilde n&\equiv& n_1+\frac{n_2}2\cos\theta_1-\frac{n_3}2\cos\theta_2.\eea
The Virasoro constraints give:
\bea &&-\frac14\kappa^2+\sin^2\xi\cos^2\xi(\tilde\omega^2+\tilde n^2)+\frac14\cos^2\xi\sin^2\theta_1(\omega_2^2+n_2^2)\nn\\
&&\,\, +\frac14\sin^2\xi\sin^2\theta_2(\omega_3^2+n_3^2)=0, \label{vir1} \eea
and
\bea \frac14(\omega_2n_2\cos^2\xi\sin^2\theta_1 +\omega_3n_3\sin^2\xi\sin^2\theta_2)+\tilde\omega\tilde n\sin^2\xi\cos^2\xi
=0. \label{vir2} \eea
The energy and the angular momentums of the string are:
\bea
E&=&\frac{\sqrt{\tilde\lambda}}4\kappa,\\
J_1&=&\sqrt{\tilde\lambda}\int\frac{d\sigma}{2\pi}\tilde\omega\sin^2\xi\cos^2\xi\nn\\
&=&\sqrt{\tilde\lambda}\tilde\omega\sin^2\xi\cos^2\xi,\\
J_2&=&\sqrt{\tilde\lambda}(\frac12\tilde\omega\sin^2\xi\cos^2\xi\cos\theta_1+\frac{\omega_2}4
\cos^2\xi\sin^2\theta_1),\\
J_3&=&\sqrt{\tilde\lambda}(-\frac12\tilde\omega\sin^2\xi\cos^2\xi\cos\theta_2+\frac{\omega_3}4
\sin^2\xi\sin^2\theta_2).\eea

\subsection{Some explicit solutions}

We now give some explicit solutions of the above equations of motion
and the Virasoro constraints.

\subsubsection{Case 1: $\xi=\pi/4, \theta_1=\theta_2=0$ or $\xi=\pi/4, \theta_1=0, \theta_2=\pi$. \label{case1}}
In this case, the equations of motion for $\theta_1, \theta_2, \xi$ are satisfied for arbitrary
$\omega_i, n_i (i=1, 2, 3)$.
The second Virasoro constraint eq.~(\ref{vir2}) gives \footnote{Here $-$ in $\mp$ corresponds to $\theta_2=0$, while
$+$ in $\mp$ corresponds to $\theta_2=\pi$. }
\be (\omega_1+\frac{\omega_2}2\mp \frac{\omega_3}2)(n_1+\frac{n_2}2\mp \frac{n_3}2)=0,\label{Vir12}\ee
While the first Virasoro constraint eq.~(\ref{vir1}) gives
\be \kappa^2= (\omega_1+\frac{\omega_2}2\mp \frac{\omega_3}2)^2+(n_1+\frac{n_2}2\mp \frac{n_3}2)^2. \label{vir11} \ee
Eq.~(\ref{Vir12}) gives either
\be n_1+\frac{n_2}2\mp \frac{n_3}2=0,\ee
or \be \omega_1+\frac{\omega_2}2\mp \frac{\omega_3}2=0.\ee
 In the first case we have:
\be J_1=2J_2=\mp 2J_3=\frac{\sqrt{\tilde\lambda}}{4}(\omega_1+\frac{\omega_2}2\mp \frac{\omega_3}2),\ee
and, using eq.~(\ref{vir11}),
\be E=\frac{\sqrt{\tilde\lambda}}{4}\kappa=|J_1|, \ee
while in the second case, we have
\be J_1=J_2=J_3=0, E=\frac{\sqrt{\tilde\lambda}}{4}\kappa=\frac{\sqrt{\tilde\lambda}}{4}|n_1+\frac{n_2}2\mp \frac{n_3}2|. \ee

\subsubsection{Case 2: $\xi=\pi/4, \theta_2=\pi-\theta_1, \omega_2=\omega_3, n_2=n_3$.\label{case2}}
\begin{figure}[ht]
    \epsfxsize=50mm%
    \hfill\epsfbox{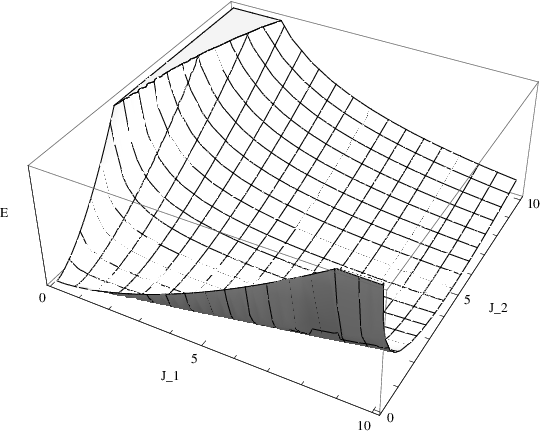}\hfill~\\
    \caption{$E$ as function of $J_1, J_2$ with $\theta=\pi/10$.}
    \label{fig1}
   \end{figure}

\begin{figure}[ht]
    \epsfxsize=50mm%
    \hfill\epsfbox{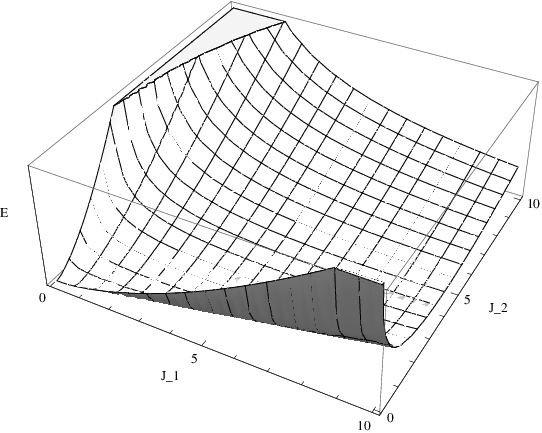}\hfill~\\
    \caption{$E$ as function of $J_1, J_2$ with $\theta=\pi/4$.}
    \label{fig2}
   \end{figure}

\begin{figure}[ht]
    \epsfxsize=50mm%
    \hfill\epsfbox{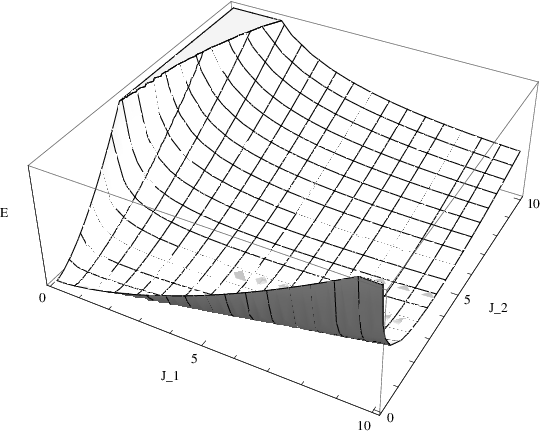}\hfill~\\
    \caption{$E$ as function of $J_1, J_2$ with $\theta=2\pi/5$.}
    \label{fig3}
   \end{figure}

Now the equation of motion for $\xi$ has been satisfied.
The equations of motion for $\theta_1, \theta_2$ give
\bea \frac{\sin2\theta_1}8(\omega_2^2-n_2^2)+\frac{\sin\theta_1}4((n_1+\cos\theta_1 n_2)
n_2-(\omega_1+\cos\theta_1\omega_2)\omega_2)=0,\label{theta1}\eea
and
\bea \frac{\sin2\theta_2}8(\omega_2^2-n_2^2)+\frac{\sin\theta_2}4((\omega_1+\cos\theta_1\omega_2)\omega_2-(n_1+\cos\theta_1 n_2)
n_2)=0,\label{theta2}\eea
respectively. Both of them lead to
\be \sin\theta_1(\omega_1\omega_2-n_1n_2)=0.\ee
If we assume that in this case $\theta_1$ is neither $0$ nor $\pi$, we get
\be \omega_1\omega_2=n_1n_2.\label{omegan}\ee
From eq.~(\ref{vir2}), we get
\be (\omega_1n_2+\omega_2n_1)\cos\theta_1+\omega_1n_1+\omega_2n_2=0, \ee
while eq.~(\ref{vir1}) gives
\bea \kappa^2&=&\omega_1^2+n_1^2+\omega_2^2+n_2^2+2\cos\theta_1(\omega_1\omega_2+n_1n_2)\nonumber\\
&=&\omega_1^2+n_1^2+\omega_2^2+n_2^2+4\cos\theta_1\omega_1\omega_2 \eea
where in the last step eq.~(\ref{omegan}) has been used.

The conserved charges of this solution are
\bea E&=&\frac{\sqrt{\tilde\lambda}}{4}\kappa=\frac{\sqrt{\tilde\lambda}}{4}\sqrt{\omega_1^2+n_1^2+\omega_2^2+n_2^2+4\cos\theta_1\omega_1\omega_2},\\
J_1&=&\frac{\sqrt{\tilde\lambda}}{4}(\omega_1+\omega_2\cos\theta_1),\\
J_2&=&\frac{\sqrt{\tilde\lambda}}{4}[\frac{\cos\theta_1}{2}(\omega_1+\omega_2\cos\theta_1)+\frac{\omega_2\sin^2\theta_1}{2}] \nn\\
&=&\frac{\sqrt{\tilde\lambda}}{8}(\omega_2+\omega_1\cos\theta_1),\label{j2}\\
J_3&=&J_2.\eea

From eqs.~(\ref{omegan})-(\ref{j2}), we can get the following relation among $E, J_1, J_2$ and $\theta_1$:
\be E=\frac{|J_1^2+4 J_2^2-4 J_1J_2\cos\theta_1|}{\sin^2\theta_1}\sqrt{\frac{\cos\theta_1}{2J_1J_2}}\ee
In Fig.~\ref{fig1}-\ref{fig3}, we plot $E$ as function of $J_1, J_2$ with $\theta=\pi/10, \pi/4, 2\pi/5$, respectively. We see that the dependence
on the value of $\theta$ is weak in some sense.


We also notice that if we start with the ansatz $\xi=\pi/4, \theta_2=\pm\theta_1, \omega_2=-\omega_3, n_2=-n_3$,
we can get the same results as above through replacing $\theta_2, \omega_3, n_3$ by $\pm\theta_1, -\omega_2, -n_2$.

\section{Multi-spin strings in $AdS_4\times CP^3_\gamma$ \label{deformed}}
\subsection{General discussions}
Now we turn to consider the $\beta$-deformed ABJM theory. We only
consider the case with real $\beta$ and rename it  $\gamma$. The
superpotential in eq.~(\ref{superpotential}) is now deformed into
\be
W_\gamma=\frac{4\pi}{k}\Tr(e^{-i\pi\gamma/2}A_1B_1A_2B_2-e^{i\pi\gamma/2}A_1B_2A_2B_1)\ee
The gravity dual of this theory was given in \cite{Imeroni:2008cr}.
For the study of classical string solutions, we only need the
background metric and the $NS-NS$ B-field \footnote{In the previous
versions of this paper, some factors are missed in the expression of
$B$-field. We would like to express our thanks to CarloAlberto Ratti
for pointing out this to us via E-mail. After fixing this, some following
calculations are corrected accordingly. And now the results here are
consisted with the result in \cite{Ratti}.}:
\be
ds^2_{\tilde\gamma}=R^2(\frac14ds^2_{AdS_4}+ds^2_{CP^3_{\tilde\gamma}}),\ee
\bea
ds^2_{CP^3_{\tilde\gamma}}&=&d\xi^2+\frac14\cos^2\xi(d\theta_1^2+G\sin^2\theta_1d\varphi_1^2)+
\frac14\sin^2\xi(d\theta_2^2\nn\\
& &\,\,+G\sin^2\theta_2d\varphi_2^2)
+G\cos^2\xi\sin^2\xi(d\psi+\frac12\cos\theta_1d\varphi_1\nn\\
& &\,\,-\frac12\cos\theta_2d\varphi_2)^2 +\tilde\gamma^2
G\sin^4\xi\cos^4\xi \sin^2\theta_1\sin^2\theta_2d\psi^2, \eea \bea
B&=&-R^2\tilde\gamma
G\sin^2\xi\cos^2\xi(\frac12\cos^2\xi\sin^2\theta_1\cos\theta_2d\psi\wedge
d\varphi_1 \nn\\&&+\frac12\sin^2\xi \sin^2\theta_2\cos\theta_1d\psi
\wedge d\varphi_2 +\frac14 fd\varphi_1\wedge d\varphi_2).\eea Here
$f$ and $G$ are defined as \be
f=\sin^2\theta_1\sin^2\theta_2+\cos^2\xi\sin^2\theta_1\cos^2\theta_2+\sin^2\xi\sin^2\theta_2\cos^2\theta_1,\ee
\be G=1/(1+\tilde\gamma^2f\sin^2\xi\cos^2\xi).\ee The relation
between $\gamma$ and $\tilde\gamma$ is \be
\tilde\gamma=\frac{R^2}{4}\gamma.\ee

Now the $\sigma$-model action is
\bea S&=&\frac{\sqrt{\tilde{\lambda}}}{4\pi}\int d\sigma d\tau \left(\frac14(\cosh^2\rho(\dot{t}^2-t^{\prime2})
-\dot{\rho}^2+\rho^{\prime2}+\sinh^2\rho(\theta^{\prime2}-\dot{\theta}^2)\right.\nn\\
&&+\sinh^2\rho\sin^2\theta(\varphi^{\prime
2}-\dot{\varphi}^2))-\dot{\xi^2}+\xi^{\prime2}+\frac14\cos^2\xi(-(\dot{\theta}_1^2
+G\sin^2\theta_1\dot{\varphi}_1^2)\nn\\
&&+\theta_1^{\prime2}+G\sin^2\theta_1\varphi^{\prime2}_1))+\frac14\sin^2\xi(-(\dot{\theta}_2^2+G\sin^2\theta_2\dot{\varphi}_2^2)
+\theta_2^{\prime2}\nn\\ &&+G\sin^2\theta_2\varphi^{\prime2}_2)
+G\sin^2\xi\cos^2\xi (-(\dot{\psi}+\frac12\cos\theta_1\dot{\varphi}_1-\frac12\cos\theta_2\dot{\varphi}_2)^2\nn\\
&&+(\psi^\prime+\frac12\cos\theta_1\varphi_1^\prime
-\frac12\cos\theta_2\varphi_2^\prime)^2)+\tilde\gamma^2G\sin^4\xi\cos^4\xi
\sin^2\theta_1\sin^2\theta_2\nn\\
&&(\psi^{\prime2}-\dot{\psi}^2)-2\tilde\gamma G\sin^2\xi\cos^2\xi(\frac12\cos^2\xi\sin^2\theta_1\cos\theta_2(\dot{\psi}\varphi_1^\prime-\psi^\prime\dot{\varphi}_1)\nn\\
&&
+\frac12\sin^2\xi\sin^2\theta_2\cos\theta_1(\dot{\psi}\varphi_2^\prime-\psi^\prime\dot{\varphi}_2)+\frac14 f(\dot{\varphi}_1\varphi_2^\prime-\varphi_1^\prime\dot{\varphi}_2))
 \left.
\right).\eea

We still use the following ansatz for the multi-spin string
solutions \be \rho=0, t=\kappa\tau, \psi=\omega_1\tau+n_1\sigma,
\varphi_1=\omega_2\tau+n_2\sigma, \varphi_2=\omega_3\tau+n_3\sigma,
\ee with $\xi, \theta_1, \theta_2$ being constants. We still have
that the equations of motion for $t, \rho, \theta, \varphi, \psi,$
$\varphi_1, \varphi_2$ have been satisfied already. Now equation of
motion for $\xi$ reads \bea && \frac12G
\sin4\xi(\tilde\omega^2-\tilde n^2)-\frac14
G\sin2\xi\sin^2\theta_1(\omega_2^2-n_2^2)+\frac14 G\sin2\xi
\sin^2\theta_2 \nn\\ && (\omega_3^2-n_3^2)+\frac{\partial
G}{\partial \xi}F+\tilde\gamma^2G\frac{\partial}{\partial
\xi}(\sin^4\xi\cos^4\xi)
\sin^2\theta_1\sin^2\theta_2(\omega_1^2-n_1^2)\nn\\&&+2\tilde\gamma
G\sin^2\xi\cos^2\xi
(-\sin\xi\cos\xi\sin^2\theta_1\cos\theta_2(\omega_1n_2-\omega_2n_1)
\nn\\ && +\sin\xi\cos\xi
\sin^2\theta_2\cos\theta_1(\omega_1n_3-\omega_3n_1)
+\frac14 \frac{\partial f}{\partial\xi}(\omega_2n_3-\omega_3n_2))\nn\\
&&+\tilde\gamma G \sin4\xi (\frac 12\cos^2\xi\sin^2\theta_1
\cos\theta_2(\omega_1n_2-\omega_2n_1)\nn\\
&&+\frac12\sin^2\xi\sin^2\theta_2\cos\theta_1(\omega_1n_3-\omega_3n_1)
+\frac14f(\omega_2n_3-\omega_3n_2)) =0,
 \eea
 where \bea F&\equiv& \frac14\cos^2\xi\sin^2\theta_1(\omega_2^2-n_2^2)+\frac14\sin^2\xi\sin^2\theta_2(\omega_3^2-n_3^2)
 \nn\\
 &&+\sin^2\xi\cos^2\xi(\tilde\omega^2-\tilde n^2) +\tilde\gamma^2\sin^4\xi\cos^4\xi\sin^2\theta_1\sin^2\theta_2(\omega_1^2-n_1^2)
 \nn\\
 &&+2\tilde\gamma\sin^2\xi\cos^2\xi(\frac14f(\omega_2n_3-\omega_3n_2)+\frac12\cos^2\xi\sin^2\theta_1
 \cos\theta_2(\omega_1n_2-\omega_2n_1)\nn\\
 &&\,\,+\frac12\sin^2\xi\sin^2\theta_2\cos\theta_1(\omega_1n_3-\omega_3n_1)).\label{defxi}
\eea

The equation of motion for $\theta_1$ is
\bea &&\frac{\partial G}{\partial \theta_1}F+\frac12G\cos^2\xi\sin\theta_1\cos\theta_1 (\omega_2^2-n_2^2)
+G\sin^2\xi\cos^2\xi\sin\theta_1(\tilde nn_2-\tilde\omega\omega_2)\nn\\
&& +2\tilde\gamma^2G\sin^4\xi\cos^4\xi\sin\theta_1\cos\theta_1\sin^2\theta_2
(\omega_1^2-n_1^2)+2\tilde\gamma G\sin^2\xi\cos^2\xi\nn\\
&&\times(\frac14\frac{\partial f}{\partial \theta_1}(\omega_2n_3-\omega_3n_2)+\cos^2\xi\sin\theta_1\cos\theta_1\cos\theta_2(\omega_1n_2-\omega_2n_1)\nn\\
&&-\frac12\sin^2\xi\sin^2\theta_2\sin\theta_1
(\omega_1n_3-\omega_3n_1))=0,  \label{deftheta1}\eea
and the equation of motion for $\theta_2$ is
\bea &&\frac{\partial G}{\partial \theta_2}F+\frac12G\sin^2\xi\sin\theta_2\cos\theta_2 (\omega_3^2-n_3^2)
+G\sin^2\xi\cos^2\xi\sin\theta_2\nn\\
&&(\tilde\omega\omega_3-\tilde nn_3)   +2\tilde\gamma^2G\sin^4\xi\cos^4\xi\sin^2\theta_1\sin\theta_2\cos\theta_2
(\omega_1^2-n_1^2)+2\tilde\gamma G\nn\\
&&\sin^2\xi\cos^2\xi(\frac14\frac{\partial f}{\partial \theta_2}(\omega_2n_3-\omega_3n_2)+\sin^2\xi\sin\theta_2\cos\theta_2\cos\theta_1(\omega_1n_3-\omega_3n_1)\nn\\
&&-\frac12\cos^2\xi\sin^2\theta_1\sin\theta_2
(\omega_1n_2-\omega_2n_1))=0. \label{deftheta2}\eea The Virasoro
constraints are \bea &&
-\frac14\kappa^2+G\sin^2\xi\cos^2\xi(\tilde\omega^2+\tilde
n^2)+\frac14G\cos^2\xi\sin^2\theta_1(\omega_2^2+n_2^2)
\nn\\
&&+\frac14
G\sin^2\xi\sin^2\theta_2(\omega_3^2+n_3^2)+\tilde\gamma^2G\sin^4\xi\cos^4\xi\sin^2\theta_1\sin^2\theta_2\nn\\
&&(\omega_1^2+n_1^2)=0,\label{defvir1}\eea
and
\bea && G\sin^2\xi\cos^2\xi\tilde\omega\tilde n+\frac14G\cos^2\xi\sin^2\theta_1\omega_2n_2+\frac14G\sin^2\xi\sin^2\theta_2
\omega_3n_3\nn\\
&&+\tilde\gamma^2G\sin^4\xi\cos^4\xi\sin^2\theta_1\sin^2\theta_2\omega_1n_1=0, \label{defvir2} \eea

The conserved charges of this solution are
\bea
E&=&\frac{\sqrt{\tilde\lambda}}4\kappa,\\
J_1&=&\sqrt{\tilde\lambda}G\int\frac{d\sigma}{2\pi}[\tilde\omega\sin^2\xi\cos^2\xi+\tilde\gamma^2\omega_1\sin^4\xi\cos^4\xi\sin^2\theta_1\sin^2\theta_2
\nn\\
&+&\tilde\gamma \sin^2\xi\cos^2\xi(\frac12n_2\cos^2\xi\sin^2\theta_1\cos\theta_2+\frac12n_3\sin^2\xi\sin^2\theta_2\cos\theta_1)]\nn\\
&=&\sqrt{\tilde\lambda}G(\tilde\omega\sin^2\xi\cos^2\xi+\tilde\gamma^2\omega_1\sin^4\xi\cos^4\xi\sin^2\theta_1\sin^2\theta_2
\nn\\
&+&\frac12\tilde\gamma
\sin^2\xi\cos^2\xi(n_2\cos^2\xi\sin^2\theta_1\cos\theta_2
\nn\\
&+&n_3\sin^2\xi\sin^2\theta_2\cos\theta_1)),\\
J_2&=&\sqrt{\tilde\lambda}G(\frac12\tilde\omega\sin^2\xi\cos^2\xi\cos\theta_1+\frac{\omega_2}4
\cos^2\xi\sin^2\theta_1\nn\\
&+& \tilde\gamma \sin^2\xi\cos^2\xi (-\frac12n_1\cos^2\xi\sin^2\theta_1\cos\theta_2+\frac14fn_3)),\\
J_3&=&\sqrt{\tilde\lambda}G(-\frac12\tilde\omega\sin^2\xi\cos^2\xi\cos\theta_2+\frac{\omega_3}4
G\sin^2\xi\sin^2\theta_2\nn\\
&+& \tilde\gamma
\sin^2\xi\cos^2\xi(-\frac12n_1\sin^2\xi\sin^2\theta_2\cos\theta_1-\frac14fn_2)).
\eea

\subsection{Some explicit solutions}

Now we give some explicit solutions. First, one can confirm by
computations that the following solution
\bea \xi&=&\frac\pi4, \theta_1=\cos^{-1}\frac{-2n_1}{n_2}, \theta_2=0,\\
\omega_1&=&\frac{2n_2}{\tilde\gamma}, \omega_2=\omega_3=0, n_3=0,
\eea satisfies the equations of motion (\ref{defxi},
\ref{deftheta1}, \ref{deftheta2}) and the second Virasoro constraint
eq.~(\ref{defvir2}).  Now we have
\bea f&=&\frac12(1-\frac{4n_1^2}{n_2^2}),\\
G&=&\left(1+\frac{\tilde\gamma^2}{8}(1-\frac{4n_1^2}{n_2^2})\right)^{-1}.
\eea The first Virasoro constraint eq.~(\ref{defvir1}) gives \bea
\kappa=\sqrt{G(\frac{\tilde\gamma^2+8}{2\tilde\gamma^2}n_2^2-2n_1^2)}=\frac{2|n_2|}{\tilde\gamma}.\eea
The conserved charges are
\bea E&=&\frac{\sqrt{\tilde\lambda}}4\kappa=\frac{\sqrt{\tilde\lambda}|n_2|}{2\tilde\gamma},\\
J_1&=&\frac{\sqrt{\tilde\lambda}}2 Gn_2(\frac{1}{\tilde\gamma}+\frac{\tilde\gamma}8(1-\frac{4n_1^2}{n_2^2}))=\frac{\sqrt{\tilde\lambda}n_2}{2\tilde\gamma},\\
J_2&=&-\frac{n_1}{n_2}J_1=-\frac{\sqrt{\tilde\lambda}n_1}{2\tilde\gamma},\\
J_3&=&-\frac12J_1.\eea
So we obtain a simple relation between $E$ and $J_1$:\be E=|J_1|.\ee

Finally we search for solutions starting with the ansatz as in subsection~\ref{case2}
\be \xi=\pi/4, \theta_2=\pi-\theta_1, \omega_2=\omega_3, n_2=n_3.\ee
Since the case when $\theta_1=0$ or $\pi$ will lead essentially to the solution discussed in subsection~\ref{case1}, below we will consider
only cases with $\sin\theta_1\neq 0$.

With non-vanishing $\tilde\gamma$,  the equations of motion for
$\xi, \theta_1, \theta_2$ lead to \be \omega_1n_2-\omega_2n_1=0,\ee
and \bea &&
-\frac{\tilde\gamma^2}4\sin\theta_1\cos\theta_1(\omega_2^2-n_2^2)+\sin\theta_1(-1-\frac12\tilde\gamma^2+\frac14\tilde\gamma^2\sin^2\theta_1)
\nn\\
&&
(\omega_1\omega_2-n_1n_2)+\sin\theta_1(\frac14\tilde\gamma^2\cos^2\theta_1-\frac12\tilde\gamma^2\cos^3\theta_1\nn\\
&&+\frac1{16}\tilde\gamma^4
\sin^4\theta_1\cos\theta_1)
(\omega_1^2-n_1^2)=0, \label{eom}\eea
while the second Virasoro constraints (eq.~(\ref{defvir2})) gives
\be n_1\omega_1+n_2\omega_2+\cos\theta_1(\omega_1n_2+\omega_2n_1)+\frac14\tilde\gamma^2\sin^4\theta_1\omega_1n_1=0. \ee

These equations give two kinds of solutions under the condition that $\sin\theta_1$ is nonzero:
\begin{itemize}

\item Solution 1 $\omega_1=\omega_2=0$: since now the string does not rotate at all, this solution is less interesting.

\item Solution 2 $n_1=n_2=0$: now the solution gives a point-like string.
\end{itemize}

For the second case, eq.~(\ref{eom}) gives
\bea && -\frac14\tilde\gamma^2\cos\theta_1\omega_2^2+(-1-\frac12\tilde\gamma^2+\frac14\tilde\gamma^2\sin^2\theta_1)
\omega_1\omega_2\nn\\
&&+(\frac14\tilde\gamma^2\cos^2\theta_1-\frac12\tilde\gamma^2\cos^3\theta_1+\frac1{16}\tilde\gamma^4\sin^4\theta_1\cos\theta_1)
\omega_1^2=0, \eea
from which we can get the expression of  $\omega_1$ in terms of other variables.

Now, the first Virasoro constraint (eq.~(\ref{defvir1})) gives
\be \kappa^2=G(\omega_1^2+\omega_2^2+2\cos\theta_1\omega_1\omega_2+\frac14\tilde\gamma^2\sin^4\theta_1\omega_1^2).\ee
The conserved change are
\bea E&=&\frac{\sqrt{\tilde\lambda}}4\sqrt{G(\omega_1^2+\omega_2^2+2\cos\theta_1\omega_1\omega_2+\frac14\tilde\gamma^2\sin^4\theta_1\omega_1^2)},\\
J_1
&=&\sqrt{\tilde\lambda}G(
\frac{\omega_1+\cos\theta_1\omega_2}4+\frac{\tilde\gamma^2\omega_1\sin^4\theta_1}{16})  ,
\nn\\
J_2&=&\frac{\sqrt{\tilde\lambda}G}8(\omega_1\cos\theta_1+\omega_2),\nn\\
J_3&=&J_2,
\eea
with \be G=\left(1+\frac{\tilde\gamma^2\sin^2\theta_1}{4}\right)^{-1}.\ee

Now we focus on the special case of $\theta=\pi/2$. Then we have
\be G=\left(1+\frac{\tilde\gamma^2}{4}\right)^{-1},\ee
and eq.~(\ref{eom}) gives \be \omega_1\omega_2=0.\ee So either $\omega_1=0$
or $\omega_2=0$. We will discuss the conversed charges in these two cases one by one:

When $\omega_1=0, \omega_2\ne 0$, we get:
\bea J_1&=&0,\nn\\
J_2&=&J_3=\frac{\sqrt{\tilde\lambda}G\omega_2}8,\nn\\
E&=&\frac{\sqrt{\tilde\lambda G}|\omega_2|}{4}. \eea
From this we get the relation among $E$ and $J_i$'s as:
\be E=2\sqrt{1+\frac{\tilde\gamma^2}4}|J_2|=2\sqrt{1+\frac{\tilde\gamma^2}4}|J_3|.\ee

When $\omega_2=0, \omega_1\ne0$, the conserved charges are:
\bea J_1&=&\sqrt{\tilde\lambda}G(\frac{\omega_1}{4}+\frac{\tilde\gamma^2\omega_1}{16})\nn\\
&=&\frac{\sqrt{\tilde\lambda}\omega_1}{4},\nn\\
J_2&=&J_3=0,\nn\\
E&=&\frac{\sqrt{\tilde\lambda}}{4}\sqrt{G(\omega_1^2+\frac14\tilde\gamma^2\omega_1^2)}=\frac{\sqrt{\tilde\lambda}|\omega_1|}{4}.  \eea
From this we get the relation among $E$ and $J_i$'s as:
\be E=|J_1|.\ee


\section{Conclusion and discussions}

In this paper, we focus on multi-spin string solutions in $AdS_4\times CP^3$ and its $\beta$-deformation.
We give explicit results for various solutions and compute their conserved charges.  Integrable structure in the $\beta$-deformed ABJM theory deserves
further studies. We expect that the spin chain of this theory could be obtained from the spin chain of the original
ABJM theory by imposing some twisted boundary conditions. This could be tested either through  direct perturbative computations
or from twisted S-matrix as the situation in  four-dimensional ${\cal N}=4$ super Yang-Mills theory \cite{Ahn:2010ws}. On the other hand, it is also interesting
to search for other type of classical strings in $AdS_4\times CP^3_\gamma$, such as folded strings. The situation here will be much more
complicated than the folded strings in $\beta$-deformed $AdS_5\times S^5$ \cite{Chen:2005sba, Chen:2006bh}. We hope to come back to these issues in the near future.

\section*{Acknowledgments}
It is my great pleasure to thank Bin Chen, Chao-Guang Huang, Song He, Daniel Louis Jafferis, Wei Li, Yi Ling, Hong Lu,  Wei Song and Tadashi
Takayanagi for very helpful discussions. Especially we would like to
express our thanks to CarloAlberto Ratti for pointing out the
mistakes in the computations of section \ref{deformed} in old
versions of the preprint. The work of JW is partly supported by NSFC
under Grant Nos. 11105154, 11222549, and by Youth Innovation
Promotion Association, CAS. JW gratefully acknowledges the support
of K.~C.~Wong Education Foundation as well.


\begin{thebibliography}{99}

\bibitem{GKP2002}
  S.~S.~Gubser, I.~R.~Klebanov and A.~M.~Polyakov,
  Nucl.\ Phys.\  B {\bf 636}, 99 (2002).

\bibitem{Frolov:2002av}
  S.~Frolov and A.~A.~Tseytlin,
  JHEP {\bf 0206}, 007 (2002)
  [hep-th/0204226].

\bibitem{Beisert:2003xu}
  N.~Beisert, J.~A.~Minahan, M.~Staudacher and K.~Zarembo,
  JHEP {\bf 0309}, 010 (2003)
  [hep-th/0306139].


\bibitem{Beisert:2003ea}
  N.~Beisert, S.~Frolov, M.~Staudacher and A.~A.~Tseytlin,
  JHEP {\bf 0310}, 037 (2003)
  [hep-th/0308117].


\bibitem{Minahan:2002ve}
  J.~A.~Minahan, K.~Zarembo,
  JHEP {\bf 0303}, 013 (2003).

\bibitem{Mal97}
J.~M.~Maldacena,
Adv.\ Theor.\ Math.\ Phys.\ {\bf 2},
231 (1998), [hep-th/9711200].

\bibitem{Gubser:1998bc}
  S.~S.~Gubser, I.~R.~Klebanov and A.~M.~Polyakov,
  Phys.\ Lett.\  B {\bf 428}, 105 (1998),
  [arXiv:hep-th/9802109].

\bibitem{Witten:1998qj}
  E.~Witten,
  Adv.\ Theor.\ Math.\ Phys.\  {\bf 2}, 253 (1998),
  [arXiv:hep-th/9802150].


\bibitem{Tseytlin:2010jv}
  A.~A.~Tseytlin,
  Lett.\ Math.\ Phys.\  {\bf 99}, 103 (2012)
  [arXiv:1012.3986 [hep-th]].

\bibitem{Kazakov:2004qf}
  V.~A.~Kazakov, A.~Marshakov, J.~A.~Minahan and K.~Zarembo,
  JHEP {\bf 0405}, 024 (2004)
  [hep-th/0402207].

\bibitem{Kazakov:2004nh}
  V.~A.~Kazakov and K.~Zarembo,
  JHEP {\bf 0410}, 060 (2004)
  [hep-th/0410105].

\bibitem{Beisert:2004ag}
  N.~Beisert, V.~A.~Kazakov and K.~Sakai,
  Commun.\ Math.\ Phys.\  {\bf 263}, 611 (2006)
  [hep-th/0410253].

\bibitem{Beisert:2005bm}
  N.~Beisert, V.~A.~Kazakov, K.~Sakai and K.~Zarembo,
  Commun.\ Math.\ Phys.\  {\bf 263}, 659 (2006)
  [hep-th/0502226].

\bibitem{SchaferNameki:2004ik}
  S.~Schafer-Nameki,
  Nucl.\ Phys.\ B {\bf 714}, 3 (2005)
  [hep-th/0412254].

\bibitem{Beisert:2005di}
  N.~Beisert, V.~A.~Kazakov, K.~Sakai and K.~Zarembo,
  JHEP {\bf 0507}, 030 (2005)
  [hep-th/0503200].

\bibitem{SchaferNameki:2010jy}
  S.~Schafer-Nameki,
  Lett.\ Math.\ Phys.\  {\bf 99}, 169 (2012)
  [arXiv:1012.3989 [hep-th]].

\bibitem{ABJM}
  O.~Aharony, O.~Bergman, D.~L.~Jafferis, J.~Maldacena,
  JHEP {\bf 0810}, 091 (2008).
 [arXiv:0806.1218 [hep-th]].


\bibitem{Minahan:2008hf}
  J.~A.~Minahan and K.~Zarembo,
  JHEP {\bf 0809}, 040 (2008),
  arXiv:0806.3951 [hep-th].

\bibitem{Bak:2008cp}
  D.~Bak and S.~-J.~Rey,
  JHEP {\bf 0810}, 053 (2008)
  [arXiv:0807.2063 [hep-th]].

\bibitem{Arutyunov:2008if}
  G.~Arutyunov, S.~Frolov,
  JHEP {\bf 0809 } (2008)  129,
  [arXiv:0806.4940 [hep-th]].

\bibitem{Stefanski:2008ik}
  B.~Stefanski, jr,
  Nucl.\ Phys.\  {\bf B808 } (2009)  80-87,
  [arXiv:0806.4948 [hep-th]].

\bibitem{arXiv:0811.1566}
  J.~Gomis, D.~Sorokin and L.~Wulff,
  JHEP\ {\bf 0903}, 015  (2009),
  [arXiv:0811.1566 [hep-th]].


\bibitem{arXiv:1009.3498}
  D.~Sorokin and L.~Wulff,
  JHEP\ {\bf 1011}, 143  (2010),
  [arXiv:1009.3498 [hep-th]].


\bibitem{arXiv:1101.3777}
  D.~Sorokin and L.~Wulff,
  arXiv:1101.3777 [hep-th].

\bibitem{Grignani:2008is}
  G.~Grignani, T.~Harmark and M.~Orselli,
  Nucl.\ Phys.\ B {\bf 810}, 115 (2009)
  [arXiv:0806.4959 [hep-th]].

\bibitem{Kristjansen:2008ib}
  C.~Kristjansen, M.~Orselli, K.~Zoubos,
  JHEP {\bf 0903}, 037 (2009),
  [arXiv:0811.2150 [hep-th]].



\bibitem{Klose:2010ki}
  T.~Klose,
   Lett.\ Math.\ Phys.\  {\bf 99}, 401 (2012).


\bibitem{Lipstein:2011}
  A.~E.~Lipstein,
  [arXiv:1105.3231 [hep-th]].

\bibitem{Chen:2008qq}
  B.~Chen, J.~-B.~Wu,
  JHEP {\bf 0809}, 096 (2008).
  [arXiv:0807.0802 [hep-th]].

\bibitem{Gromov:2008bz}
  N.~Gromov and P.~Vieira,
  JHEP {\bf 0902}, 040 (2009)
  [arXiv:0807.0437 [hep-th]].

\bibitem{Astolfi:2008ji}
  D.~Astolfi, V.~G.~M.~Puletti, G.~Grignani, T.~Harmark and M.~Orselli,
  Nucl.\ Phys.\ B {\bf 810}, 150 (2009)
  [arXiv:0807.1527 [hep-th]].

\bibitem{McLoughlin:2008ms}
  T.~McLoughlin and R.~Roiban,
  JHEP {\bf 0812}, 101 (2008)
  [arXiv:0807.3965 [hep-th]].

\bibitem{Alday:2008ut}
  L.~F.~Alday, G.~Arutyunov and D.~Bykov,
  JHEP {\bf 0811}, 089 (2008)
  [arXiv:0807.4400 [hep-th]].

\bibitem{Krishnan:2008zs}
  C.~Krishnan,
  JHEP {\bf 0809}, 092 (2008)
  [arXiv:0807.4561 [hep-th]].

\bibitem{McLoughlin:2008he}
  T.~McLoughlin, R.~Roiban and A.~A.~Tseytlin,
  JHEP {\bf 0811}, 069 (2008)
  [arXiv:0809.4038 [hep-th]].

\bibitem{Bandres:2009kw}
  M.~A.~Bandres and A.~E.~Lipstein,
  JHEP {\bf 1004}, 059 (2010)
  [arXiv:0911.4061 [hep-th]].

\bibitem{Bombardelli:2008qd}
  D.~Bombardelli and D.~Fioravanti,
  JHEP {\bf 0907}, 034 (2009)
  [arXiv:0810.0704 [hep-th]].

\bibitem{Astolfi:2009qh}
  D.~Astolfi, V.~G.~M.~Puletti, G.~Grignani, T.~Harmark and M.~Orselli,
  JHEP {\bf 1004}, 079 (2010)
  [arXiv:0912.2257 [hep-th]].

\bibitem{Abbott:2010yb}
  M.~C.~Abbott, I.~Aniceto and D.~Bombardelli,
  JHEP {\bf 1012}, 040 (2010)
  [arXiv:1006.2174 [hep-th]].


\bibitem{Astolfi:2011ju}
  D.~Astolfi, V.~G.~M.~Puletti, G.~Grignani, T.~Harmark and M.~Orselli,
  JHEP {\bf 1105}, 128 (2011)
  [arXiv:1101.0004 [hep-th]].

\bibitem{Abbott:2011tp}
  M.~C.~Abbott, I.~Aniceto and D.~Bombardelli,
  J.\ Phys.\ A {\bf 45}, 335401 (2012)
  [arXiv:1111.2839 [hep-th]].

\bibitem{Astolfi:2011bg}
  D.~Astolfi, G.~Grignani, E.~Ser-Giacomi and A.~V.~Zayakin,
  JHEP {\bf 1204}, 005 (2012)
  [arXiv:1111.6628 [hep-th]].

\bibitem{Beccaria:2012qd}
  M.~Beccaria, G.~Macorini, C.~Ratti and S.~Valatka,
  JHEP {\bf 1205}, 030 (2012)
  [Erratum-ibid.\  {\bf 1205}, 137 (2012)]
  [arXiv:1203.3852 [hep-th]].

\bibitem{Beccaria:2012vb}
  M.~Beccaria, G.~Macorini, C.~Ratti and S.~Valatka,
  arXiv:1209.3205 [hep-th].
\bibitem{Wu:2011rj}
  J.~-B.~Wu,
  arXiv:1111.3727 [hep-th].

\bibitem{Leigh:1995ep}
  R.~G.~Leigh and M.~J.~Strassler,
  Nucl.\ Phys.\ B {\bf 447}, 95 (1995)
  [hep-th/9503121].


\bibitem{Lunin:2005jy}
  O.~Lunin and J.~M.~Maldacena,
  JHEP {\bf 0505}, 033 (2005)
  [hep-th/0502086].


\bibitem{Khoze:2005nd}
  V.~V.~Khoze,
  JHEP {\bf 0602}, 040 (2006)
  [hep-th/0512194].

\bibitem{Gao:2006mw}
  P.~Gao and J.~-B.~Wu,
  Nucl.\ Phys.\ B {\bf 798}, 184 (2008)
  [hep-th/0611128].

\bibitem{Witten:2003nn}
  E.~Witten,
  Commun.\ Math.\ Phys.\  {\bf 252}, 189 (2004)
  [hep-th/0312171].

\bibitem{Kulaxizi:2004pa}
  M.~Kulaxizi and K.~Zoubos,
  Nucl.\ Phys.\ B {\bf 738}, 317 (2006)
  [hep-th/0410122].


\bibitem{Roiban:2003dw}
  R.~Roiban,
  JHEP {\bf 0409}, 023 (2004)
  [hep-th/0312218].

\bibitem{Berenstein:2004ys}
  D.~Berenstein and S.~A.~Cherkis,
  Nucl.\ Phys.\ B {\bf 702}, 49 (2004)
  [hep-th/0405215].

\bibitem{Beisert:2005if}
  N.~Beisert and R.~Roiban,
  JHEP {\bf 0508}, 039 (2005)
  [hep-th/0505187].

\bibitem{Arutyunov:2010gu}
  G.~Arutyunov, M.~de Leeuw and S.~J.~van Tongeren,
  JHEP {\bf 1102}, 025 (2011)
  [arXiv:1009.4118 [hep-th]].

\bibitem{Ahn:2010ws}
  C.~Ahn, Z.~Bajnok, D.~Bombardelli and R.~I.~Nepomechie,
  JHEP {\bf 1102}, 027 (2011)
  [arXiv:1010.3229 [hep-th]].

\bibitem{Frolov:2005ty}
  S.~A.~Frolov, R.~Roiban and A.~A.~Tseytlin,
  JHEP {\bf 0507}, 045 (2005)
  [hep-th/0503192].

\bibitem{Chen:2005sba}
  H.~-Y.~Chen and S.~P.~Kumar,
  JHEP {\bf 0603}, 051 (2006)
  [hep-th/0511164].

\bibitem{Chen:2006bh}
  H.~-Y.~Chen and K.~Okamura,
  JHEP {\bf 0602}, 054 (2006)
  [hep-th/0601109].


\bibitem{Bobev:2005cz}
  N.~P.~Bobev, H.~Dimov and R.~C.~Rashkov,
  Bulg.\ J.\ Phys.\  {\bf 35}, 274 (2008)
  [hep-th/0506063].

\bibitem{Ryang:2005pg}
  S.~Ryang,
  JHEP {\bf 0511}, 006 (2005)
  [hep-th/0509195].

\bibitem{Chu:2006ae}
  C.~-S.~Chu, G.~Georgiou and V.~V.~Khoze,
  JHEP {\bf 0611}, 093 (2006)
  [hep-th/0606220].

\bibitem{Bobev:2007bm}
  N.~P.~Bobev and R.~C.~Rashkov,
  Phys.\ Rev.\ D {\bf 76}, 046008 (2007)
  [arXiv:0706.0442 [hep-th]].

\bibitem{Bykov:2008bj}
  D.~V.~Bykov and S.~Frolov,
  JHEP {\bf 0807}, 071 (2008)
  [arXiv:0805.1070 [hep-th]].

\bibitem{Ahn:2010da}
  C.~Ahn and P.~Bozhilov,
  JHEP {\bf 1007}, 048 (2010)
  [arXiv:1005.2508 [hep-th]].

\bibitem{Giardino:2011jy}
  S.~Giardino and V.~O.~Rivelles,
  JHEP {\bf 1107}, 057 (2011)
  [arXiv:1105.1353 [hep-th]].

\bibitem{Panigrahi:2012bm}
  K.~L.~Panigrahi, P.~M.~Pradhan and P.~K.~Swain,
  JHEP {\bf 1206}, 057 (2012)
  [arXiv:1203.3057 [hep-th]].


\bibitem{Ahn:2012hs}
  C.~Ahn, D.~Bombardelli and M.~Kim,
  Phys.\ Lett.\ B {\bf 710}, 467 (2012)
  [arXiv:1201.2635 [hep-th]].

\bibitem{Zoubos:2010kh}
  K.~Zoubos,
  Lett.\ Math.\ Phys.\  {\bf 99}, 375 (2012)
  [arXiv:1012.3998 [hep-th]].

\bibitem{Imeroni:2008cr}
  E.~Imeroni,
  JHEP {\bf 0810}, 026 (2008)
  [arXiv:0808.1271 [hep-th]].


\bibitem{Schimpf:2009rk}
  M.~Schimpf and R.~C.~Rashkov,
  Mod.\ Phys.\ Lett.\ A {\bf 24}, 3227 (2009)
  [arXiv:0908.2246 [hep-th]].

\bibitem{Ahn:2011nm}
  C.~Ahn and P.~Bozhilov,
  Phys.\ Lett.\ B {\bf 703}, 186 (2011)
  [arXiv:1106.3686 [hep-th]].

\bibitem{Grignani:2008te}
  G.~Grignani, T.~Harmark, M.~Orselli and G.~W.~Semenoff,
  JHEP {\bf 0812}, 008 (2008)
  [arXiv:0807.0205 [hep-th]].

\bibitem{BL1}
J.~Bagger and N.~Lambert, 
Phys. Rev. D{\bf 75},045020(2007)[arXiv:hep-th/0611108].

\bibitem{BL2}
J.~Bagger and N.~Lambert,
Phys. Rev. D{\bf 77}, 065008(2008) [arXiv:0711.0955].

\bibitem{BL3}
J.~Bagger and N.~Lambert, 
JHEP {\bf 0802}, 105(2008)[arXiv:0712.3738].

\bibitem{G1}A. Gustavsson, 
Nucl.\ Phys.\ B {\bf 811}, 66 (2009), [arXiv:0709.1260].

\bibitem{G2}A. Gustavsson, 
Nucl.\ Phys.\ B {\bf 807}, 315 (2009), [arXiv:0805.4443].

\bibitem{Berman:2008be}
  D.~S.~Berman, L.~C.~Tadrowski and D.~C.~Thompson,
  Nucl.\ Phys.\ B {\bf 802}, 106 (2008)
  [arXiv:0803.3611 [hep-th]].

\bibitem{Akerblom:2009gx}
  N.~Akerblom, C.~Saemann and M.~Wolf,
  Nucl.\ Phys.\ B {\bf 826}, 456 (2010)
  [arXiv:0906.1705 [hep-th]].

\bibitem{Berman:2007tf}
  D.~S.~Berman and L.~C.~Tadrowski,
  Nucl.\ Phys.\ B {\bf 795}, 201 (2008)
  [arXiv:0709.3059 [hep-th]].

\bibitem{Frolov:2003qc}
  S.~Frolov and A.~A.~Tseytlin,
  Nucl.\ Phys.\ B {\bf 668}, 77 (2003)
  [hep-th/0304255].

\bibitem{Giardino:2011dz}
  S.~Giardino and H.~L.~Carrion,
  JHEP {\bf 1108}, 057 (2011)
  [arXiv:1106.5684 [hep-th]].

\bibitem{Giardino:2011uc}
  S.~Giardino,
  JHEP {\bf 1112}, 022 (2011)
  [arXiv:1110.3682 [hep-th]].

\bibitem{Ratti}
C.~Ratti, [arXiv: 1211.4694[hep-th]].

\end{thebibliography}
\end{document}